\documentclass[10pt, conference]{IEEEtran}
\IEEEoverridecommandlockouts
\usepackage{cite}

\usepackage{array} 
\usepackage{amsmath,amssymb,amsfonts}
\usepackage{algorithmic}
\usepackage{graphicx}
\usepackage{textcomp}
\usepackage{xcolor}
\usepackage[numbers]{natbib}
\usepackage{graphicx}
\usepackage{adjustbox}
\usepackage{booktabs}
\usepackage{caption} 
\usepackage{multirow}
\usepackage{rotating}
\usepackage{lipsum}
\usepackage[capitalise]{cleveref}
\usepackage{comment}

\def\BibTeX{{\rm B\kern-.05em{\sc i\kern-.025em b}\kern-.08em
    T\kern-.1667em\lower.7ex\hbox{E}\kern-.125emX}}
\begin{document}


\title{\huge Towards a Signal Detection Based Measure for Assessing\\ Information Quality of Explainable Recommender Systems}


\author{\IEEEauthorblockN{Yeonbin Son}
\IEEEauthorblockA{\textit{Systems and Information Engineering} \\
\textit{University of Virginia}\\
Charlottesville, United States \\
ybson@virginia.edu}
\and
\IEEEauthorblockN{Matthew L. Bolton, \textit{Senior Member, IEEE}}
\IEEEauthorblockA{\textit{Systems and Information Engineering} \\
\textit{University of Virginia}\\
Charlottesville, United States \\
matthewlbolton@virginia.edu}
}

\maketitle

\begin{abstract}
There is growing interest in explainable recommender systems that provide recommendations along with explanations for the reasoning behind them. 
When evaluating recommender systems, most studies focus on overall recommendation performance. Only a few assess the quality of the explanations. 
Explanation quality is often evaluated through user studies that subjectively gather users' opinions on representative explanatory factors that shape end-user's perspective towards the results, not about the explanation contents itself. We aim to fill this gap by developing an objective metric to evaluate \textit{Veracity}: the information quality of explanations. 
Specifically, we decompose \textit{Veracity} into two dimensions: \textit{Fidelity} and \textit{Attunement}. 
\textit{Fidelity} refers to whether the explanation includes accurate information about the recommended item.
\textit{Attunement} evaluates whether the explanation reflects the target user’s preferences. 
By applying signal detection theory, we first determine decision outcomes for each dimension and then combine them to calculate 
 a sensitivity, which serves as the final Veracity value.
To assess the effectiveness of the proposed metric, we set up four cases with varying levels of information quality to validate whether our metric can accurately capture differences in quality.
The results provided meaningful insights into the effectiveness of our proposed metric.
\end{abstract}

\begin{IEEEkeywords}
Explainable recommender systems,
information quality,
signal detection theory,
veracity,
fidelity,
attunement
\end{IEEEkeywords}
 
\section{Introduction} \label{sec:intro}
With the rapid development of recommender systems, research has been attempting to develop methods for explaining recommendation results \cite{zhang_explainable_2020}.
Explainable recommender systems (XRS) not only provide recommendation results, but also explain why the results were generated \cite{zhang_explainable_2020}.
The purpose is to enhance users’ trust in the system by offering transparent and scrutable results to  users \cite{ghazimatin_elixir_2021}.
Such approaches typically utilize users’ behavior and the characteristics of items. 
The way of explaining recommendation results varies by the type of data that a recommender system used.
When evaluating the performance of XRS, most studies focus on examining recommendation performance.
The most representative metric for recommendation performance is accuracy, which measures the ratio of user-preferred items among the recommended item list, where a user-preferred item is one that a user either rated 4 or higher, or previously purchased by the user.
Another commonly used metric is root mean squared error (RMSE) \cite{son_improving_2020}, which is applied when the problem is rating prediction.
This metric compares pre-rated and predicted rating values to calculate the difference.

\begin{table}[]
\centering
\parbox{2.96in}{
\caption{The Seven Explanatory Factors From \cite{aggarwal_recommender_2016}.}\label{tab:factors}}\vspace{-5pt}
\begin{tabular}{@{}ll@{}}
\toprule
Explanatory criteria & Definition                                 \\ \midrule
Transparency         & Explain how the system works               \\
Efficiency           & Help users make decisions faster           \\
Trust                & Increase users’ confidence in the system   \\
Satisfaction         & Increase the ease of use or enjoyment      \\
Persuasiveness       & Convince users to try or buy               \\
Effectiveness        & Help users make good decisions             \\
Scrutability         & Allow users to tell the system it is wrong \\ \bottomrule
\end{tabular}
\vspace{-12pt}
\end{table}
    
XRS research evaluates explanations as well, but there are rare cases where the objective information quality of the explanations is measured.
General evaluation methods are divided into offline methods, online methods, and user studies.
In offline methods, the assumption is that not every recommendation case has an explanation \cite{abdollahi_explainable_2016}.
Only few of the recommendations can explain the reason, so the quality is calculated by checking whether each recommendation has explanations or not.
In online methods, the underlying assumption is that researchers can track end-user behavior after receiving recommendation results \cite{moller_explaining_2024}.
The researchers compare the user's behavior with and without explanations to determine if their is a valid difference.
For user studies, researchers interview or poll end-users to assess their satisfaction with the explanations based on seven representative explanatory factors, as shown in \cref{tab:factors} \cite{zhang_explainable_2020}. 
Each explanation is optionally rated on these factors using psychometric scales such as Likert scale.

\begin{figure*}[]
    \centering

    \includegraphics[height=0.37\textwidth]{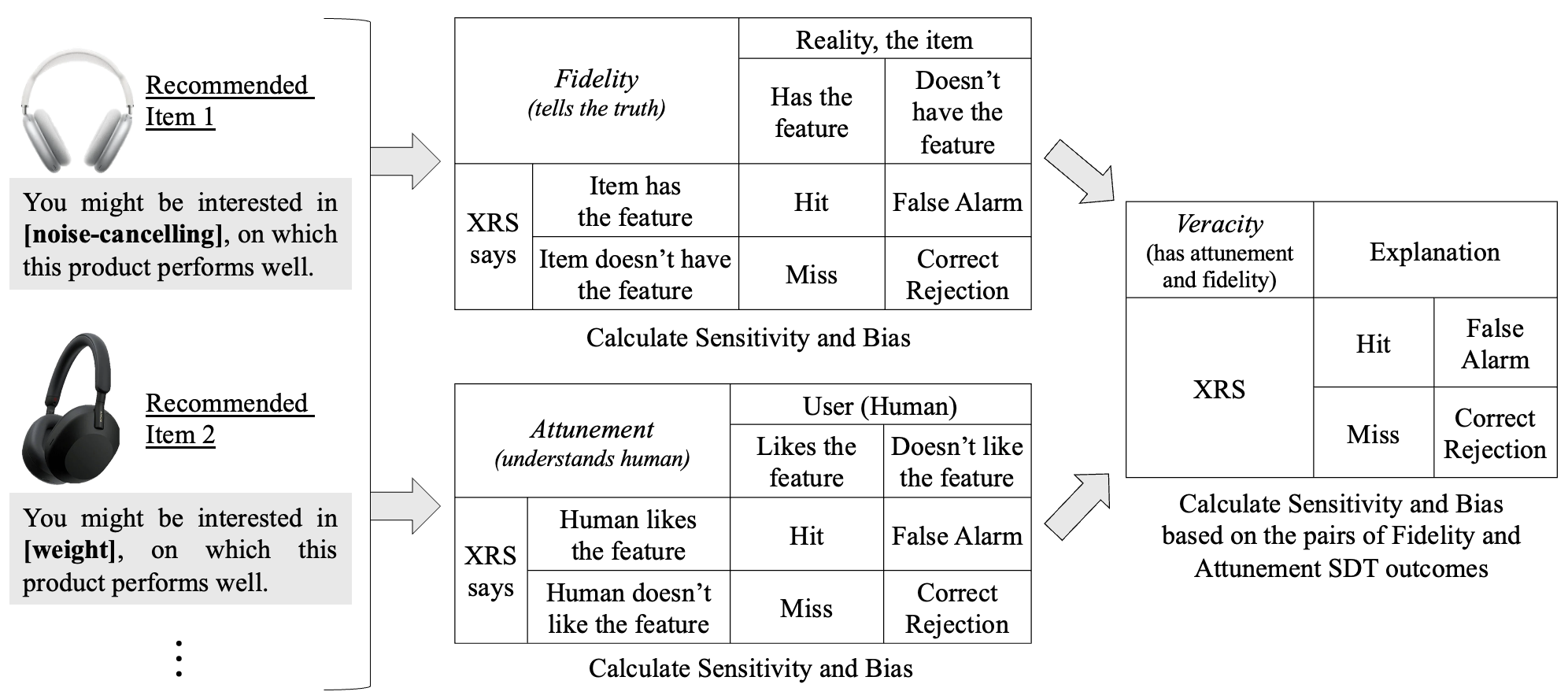}
    \caption{Illustration of our approach to assessing explanation versatility in terms of Fidelity and Attunement. First, recommendation–explanation pairs are generated for a target user. Next, Fidelity and Attunement SDT outcomes are measured based on the reality of information contained about the product (Fidelity) and the user’s feedback about feature preferences (Attunement). Veracity is calculated from the paired fidelity and attunement SDT outcomes. Sensitivity ($A'$) and bias ($B_D''$) metrics can be computed for all three measured concepts: fidelity, attunment, and veracity.}
    \label{fig:overview}
    \vspace{-10pt}
\end{figure*}

While user studies provide subjective assessments, these factors do not \textit{directly} account for the information quality.
This is a critical dimension because modern systems, including but not limited to large language models, can potentially create explanations that are not rooted in reality, are highly stochastic, or based on imperfect information. 
This led us to consider: what if there were an objective measure to evaluate the quality of explanations, particularly in terms of information quality?
Thus, the goal for this paper was to develop a metric to assess explanation quality, specifically focusing on the quality/truthfulness of provided information content, a concept we call \textit{Veracity}.
Because recommender explanations make statements that concern the truthfulness of both the product being recommended (e.g., this item has a given feature) and the person the recommendation is targeted to (e.g., you like things with the feature), Veracity needs to assess the quality of both. 
Thus, we divide Veracity into two sub-factors.
The first, \textit{Fidelity}, evaluates whether the explanation contains accurate information related to the recommended item.
The second, \textit{Attunement}, assesses whether the explanation effectively captures the user’s actual preferences.

To measure these sub-factors and combine them, we utilize signal detection theory (SDT).
SDT is a method that quantifies a decision maker’s ability to detect the presence of a phenomenon. 
SDT's critical feature is its ability to distinguish between the decision maker's sensitivity (their ability to differentiate between signal and noise/uncertainty) and the criteria it uses to make decision (its response bias) \cite{peterson_theory_1954}.
This paper explores how SDT can be used to assess explanation Veracity. This is accomplished by using SDT to quantify explanation Fidelity and Attunement separately. Then, these are combined together into a proper composite SDT measure of Veracity. 
The overview of our concept is described in Figure \ref{fig:overview}.

In what follows, we provide background for understanding our approach to measuring explanation Veracity. 
This is followed by the details of our SDT-based approach. 
We then present an experiment to evaluate different formulations of our approach
and discuss their significance. 
% ...

\section{Background}
\subsection{Explainable Recommender Systems}
XRS is an extension of traditional recommender systems designed to BOTH provide recommendations and explain why a particular item is recommended to the end-user.
Depending on the way explanations are generated, methods are categorized into model-dependent and model-independent approaches.
The model-dependent approach relies on the specific structure and internal mechanisms of the recommendation algorithm to generate explanations.
It is typically applied to collaborative filtering or content-based recommendation algorithms.
In contrast, the model-independent approach generates explanations independently of the underlying model, focusing on analyzing the recommendation output rather than the algorithm itself.
Explanations are often produced through a post-hoc analysis of recommendation results. The examples of the representative methods are local interpretable model-agnostic explanations (LIME) \cite{ribeiro2016lime} or shapley additive explanations (SHAP) \cite{lundberg2017shap}.

There are different types of explanations, and the type is heavily influenced by the nature of the utilized data.
One common method involves generating explanations that emphasize an item’s specific characteristics that a target user may like.
Here, the characteristics can be explicit (e.g. genre, actor, production company, etc. in the case of a movie recommendation) or implicit, (e.g. aspects extracted based on machine learning methods.)
Another common type is sentence-based explanation.
Such explanations can be template-based or generation-based.
In template-based explanations, a defined template is used, and the sentence is completed by inserting user-targeted words.
For generation-based explanations, natural language generation methods, such as gate recurrent networks (GRNs) \cite{cho2014learning}, transformers \cite{vaswani2017attention}, or generative pre-trained transformers (GPTs) \cite{brown2020language}, are employed to create sentences that explain the reasoning behind a recommendation.
Additionally, there are methods create explanations through images.
Some of these presents the entire image as an explanation, while others highlight specific regions of interest.

To evaluate the performance of XRS, researchers typically conduct two types of experiments: one to assess recommendation performance and the other to evaluate the quality of explanations.
In this section, we focus on explanation quality assessment, which is typically divided into three categories: offline, online, and user study.
In offline methods, mean explainability precision (MEP) and mean explainability recall (MER) are commonly calculated.
EP represents the proportion of purchased items included in the explanations relative to the number of recommended items. ER indicates the proportion of purchased items mentioned in the explanations relative to the total number of recommended items included in the explanations \cite{abdollahi_explainable_2016}.
These two metrics focus on whether explanations can be generated but do not address the quality of the generated content.
When machine-generated sentences are used as explanations, metrics such as BLEU \cite{papineni_bleu_2001} or ROUGE \cite{lin_rouge_2004} scores are calculated to evaluate the quality of the generated sentences. 
However, these do not consider users' satisfaction with the system results.
For online methods, if there is sufficient infrastructure to provide recommendations and explanations while tracking users' subsequent behavior, it is possible to calculate metrics such as conversion or click-through rates within the system \cite{moller_explaining_2024}.
However, clicking does not necessarily mean the user is satisfied with the system results.

Another approach is to conduct user studies to evaluate people's perceptions of the system's results \cite{aggarwal_recommender_2016}.
As we discussed in Section \ref{sec:intro}, seven representative factors described in Table \ref{tab:factors} are optionally assessed. These are usually measured subjectively.
Transparency ensures that users can gain a clear understanding of the system's underlying processes, making it easier to trust its functionality.
Efficiency focuses on helping users arrive at decisions more quickly by streamlining the process and eliminating unnecessary complexity.
To build trust, the system should provide consistent and accurate results that users can rely on with confidence.
A satisfying experience arises when the system is intuitive and enjoyable, encouraging continued use.
Persuasiveness plays a role in attracting users, encouraging them to explore the system further or take desired actions, such as making a purchase.
For effectiveness, the system must guide users toward making sound and beneficial decisions.
Lastly, scrutability allows users to engage with the system on a deeper level by pointing out errors or discrepancies, enabling improvement.

\subsection{Signal Detection Theory} \label{sec:sdt}
\begin{figure*}[]
    \centering
    \includegraphics[height=0.49\textwidth]{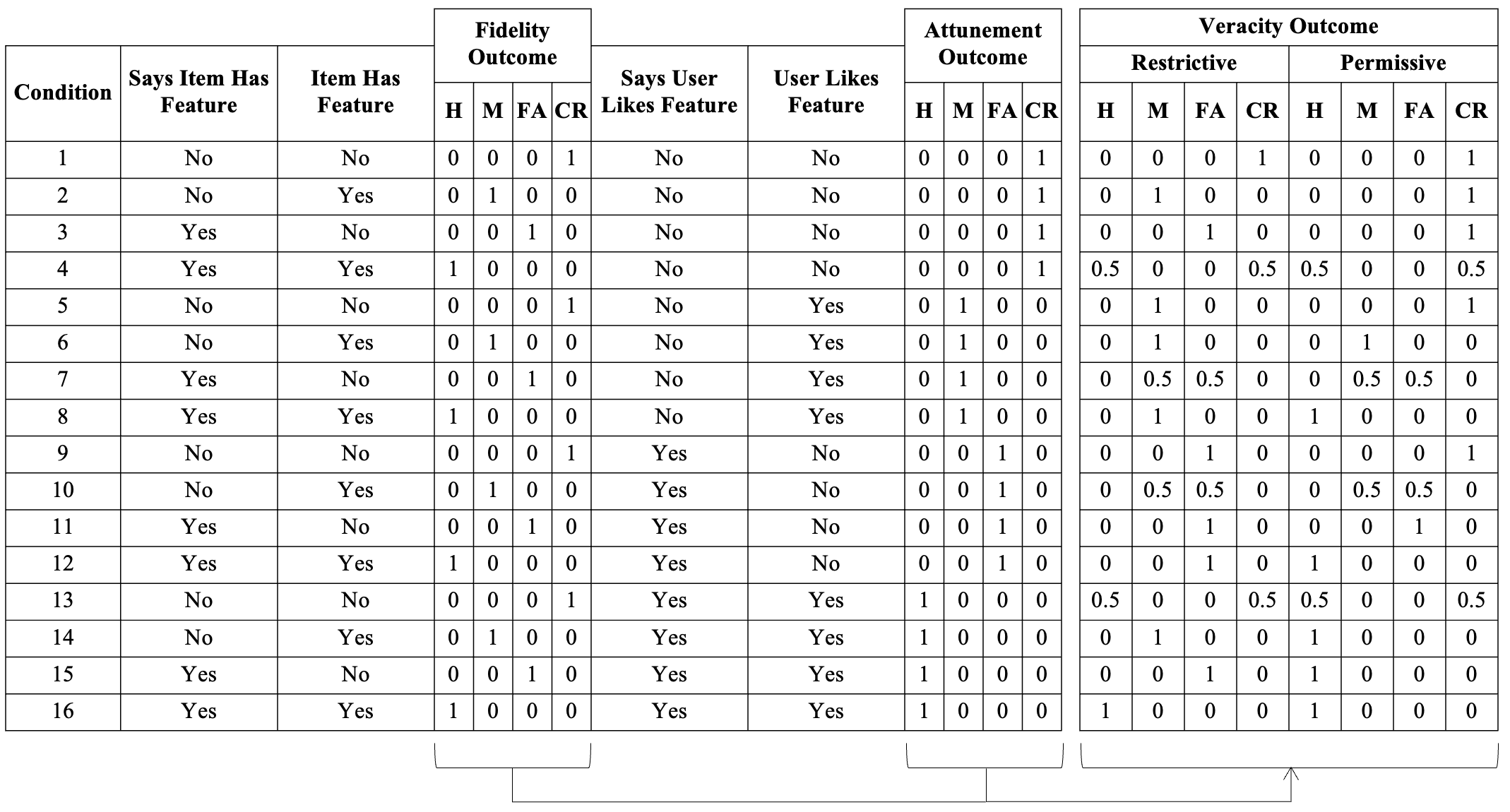}
    \caption{Illustration of the different Fidelity, Attunement, and Veracity outcomes associated with an explanation statement that claims an item has a feature that a user likes. Veracity outcomes for each condition are derived from those for Fidelity and Attunement in two different ways: a Restrictive one that punishes incorrect outcomes (M or FA) and a Permissive one that rewards correct ones (H, CR) when they are mixed between those from Fidelity and Attunement. In all of these, a value from 0 to 1 is used to indicate the extent to which a given condition produces the associated outcome. In conditions where outcomes are less than 1 (e.g. 0.5), the sum of all outcome values for a given condition will sum to 1.}
    \label{fig:cases}
    \vspace{-10pt}

\end{figure*}

SDT is a theoretical framework that shows how individuals or decision processes distinguish signals (meaningful targets) from noise (distractions that interrupt signal) under uncertainty \cite{peterson_theory_1954}.
As stated previously, this theory provides a framework to assess the sensitivity of detection (how well the decision maker can separate signal from noise) and the decision-making criteria (called bias; the strength of signal + noise that results in the judge saying yes) under uncertain conditions as separate phenomena.
Both are determined by the different decision outcomes. 
If signal is judged as present, a hit occurs. 
If noise is recognized as a signal, the outcome is a false alarm (FA). If a signal is recognized as noise, the outcome is a miss.
Finally, if noise is judged as noise, a correct rejection (CR) occurs.

Sensitivity is calculated based on the rates of outcomes. 
Hit rate (HR) represents the proportion of signal trials where the signal is detected correctly, as shown in Equation \ref{eq:hr}.
\begin{equation}
HR = \frac{{\text{Number of Hits}}} {{\text{Number of Signal Trials}}}
\label{eq:hr}
\end{equation}
False alarm rate (FAR) is calculated by Equation \ref{eq:far}.
\begin{equation}
FAR = \frac{{\text{Number of False Alarms}}} {{\text{Number of Noise Trials}}}
\label{eq:far}
\end{equation}
Note that these definitions imply that HR = 1 - miss rate (MR) and FAR = 1 - CR rate. Thus, convention is to work exclusively with HR and FAR to minimize the number of variables. 
Using HR and FAR, nonparametric sensitivity $A'$ is calculated by Equation \ref{eq:d}.
\begin{equation}
A' =
\begin{cases}
0.5 + \frac{(HR - FAR)(1 + HR - FAR)}{4HR(1 - FAR)} & \text{if } HR \geq FAR, \\
0.5 + \frac{(FAR - HR)(1 + FAR - HR)}{4FAR(1-HR)} & \text{otherwise.}
\end{cases}
\label{eq:d}
\end{equation}
$A'$ values vary from 0.5 (no discrimination between signal and noise) to 1 (perfect discrimination).

Bias provides insight into whether a person has a tendency to overreport or underreport signals. This is calculated nonparametrically by Equation \ref{eq:bias}.
\begin{equation}
B''_D = \frac{(1 - HR)(1 - FAR) -HR \cdot FAR}{(1 - HR)(1 - FAR) + HR \cdot FAR}
\label{eq:bias}
\end{equation}
A $B''_D=0$  means there is no bias, and thus we can say the user is neutral.
If $B''_D>0$, we say there is conservative bias because the user tends to report ``noise'' more often.
If $B''_D<0$, there is a liberal bias. This means the judge tends to report ``signal'' more often.

Signal detection theory has been used successfully in a number of different domains to evaluate the detection capabilities of humans \cite{gawronski_signal-detection_2024}, medical tests \cite{Swets1988, McFall2009}, and automated processes \cite{Meyer2014, Zhou2022}. 

\section{Method}
\label{sec:method}
Our approach for objectively evaluating XRS explanation Veracity starts by using SDT to separately assess produced explanations based on Fidelity and Attunement. It then combines the results into a composite SDT analysis. 
Practically, any statement in an explanation
can be assessed for Fidelity by determining the SDT outcome (H, M, FA, CR) with respect to the truthful of a given statement about the object being recommended. 
Similarly, the Attunement SDT outcome of any statement can be assessed based on its truthfulness about the preferences of a user. 
For example, a common type of explanation will contain statements of the form ``this product has X feature, which you may like.'' This means that SDT outcomes can be generated based on the truthfulness of ``this product has X feature'' claim (Fidelity) and the ``you may like this feature'' claim (Attunement).
Fig. \ref{fig:cases} illustrates all 16 of the different combinations/conditions of  Fidelity and Attunement outcomes that can be associated with explanations of this form. 



Next, the two decision outcomes are combined to determine Veracity's decision outcome. 
There are multiple ways this could be done. 
In this work, we explore two: a restrictive one and a permissive one.
Both of these assume that if the the Fidelity and Attunement outcomes match, then the Veracity outcomes matches that outcome. Similarly, if the Fidelity and Attunement outcomes both indicate correctness (with both a H and a CR) or incorrectness (both a M and a FA), then both outcomes are treated as half (0.5) occurring. 
The difference between the restrictive and permissive approaches comes when there is a discrepancy between the correctness of the Fidelity and Attunement outcomes: when one indicates a correct part of an explanation (H or CR) and the other does not (M or FA). 
In this situation, the restrictive approach gives full weight to the M or FA outcome, while the permissive approach gives it to the H or CR one. 
This process is illustrated across the conditions in Fig. \ref{fig:cases}. 
Either approach could potentially be useful. The permissive one could be helpful if the analysts think that the decision making/purchasing scenario is not ``high stakes'' or if they think that users will be forgiving of recommendations that are only partially true. Conversely, the restrictive approach could be more appropriate in scenarios where the purchasing decisions are non-trivial and/or in situations where there could be repercussions (i.e., upset users, returned items, customer loss) for even partially incorrect recommendations.

After determining the outcome of the selection decision for all pairs of recommendations, we use that results to create a confusion matrix and calculate the sensitivity and bias as described in Section \ref{sec:sdt} (via \cref{eq:d} and \cref{eq:bias}, respectively) based on the associated outcome rates.
To support diagnosticity in analyses, this is done for all three dimensions: Fidelity, Attunement, and Veracity.

This approach is novel and there is scant literature on the most appropriate way of synthesizing multiple SDT outcomes into a single SDT system.
Thus, it was unclear how the different SDT measures, including the restrictive and permissive versions of Veracity, would perform with real explanations. 
To conducted an experiment to explore this.

\section{Experimental Evaluation}
\label{sec:exp}
We conducted a simple experiment to evaluate our explanation Veracity metric. 
This experiment had two goals. 
The first was to assess how well it differentiated between different quality levels of explanations. 
The second was to compare the two conditions (restrictive vs. permissive) used by the metric to see how it impacted the differentiation. 
The following sections describe this experiment and present its results.

\subsection{Experimental Design}

The dataset we used for our experiments is MovieLens 1M, which was created by UMN \cite{movielens1m}. This is one of the most widely used dataset in recommender system research.
The dataset consists of user profiles, movie profiles, and user-movie ratings.
There are 6,040 user profiles, and their attributes include age, gender, and occupation.
The number of movie profiles is 3,706, and their attributes include actor, category, cinematographer, composer, director, editor, producer, and production company.
User ratings of the movies range from 1 to 5 with a 1-point interval, totaling 1,000,209 ratings.
Based on this dataset, we got the knowledge graph generated in \cite{cao_unifying_2019}.
The baseline explanations (with the highest quality) are generated using the state-of-the-art explainable recommendation method proposed by \cite{balloccu_post_2022}.

\begin{table*}[h]
\centering
\caption{Fidelity, Attunement, and Veracity sensitivity values ($A'$) and bias ($B''_D$) by different levels of explanation quality.}\vspace{-5pt}
\begin{tabular}{@{}cp{8cm}rrrrrrrr@{}}
\toprule
\multirow{3}{*}{Case} & \multirow{3}{*}{Description}                                                         & \multicolumn{2}{c}{\multirow{2}{*}{Fidelity}} & \multicolumn{2}{c}{\multirow{2}{*}{Attunement}} & \multicolumn{4}{c}{Veracity}                                            \\ \cmidrule(l){7-10} 
                      &                                                                                      & \multicolumn{2}{c}{}                          & \multicolumn{2}{c}{}                            & \multicolumn{2}{c}{Restrictive}    & \multicolumn{2}{c}{Permissive}     \\ \cmidrule(l){3-4} \cmidrule(l){5-6} \cmidrule(l){7-10} 
                      &                                                                                      & \multicolumn{1}{l}{$A'$}       & $B''_D$      & \multicolumn{1}{l}{$A'$}        & $B''_D$       & \multicolumn{1}{l}{$A'$} & $B''_D$ & \multicolumn{1}{l}{$A'$} & $B''_D$ \\ \midrule
1                     & Random features used as explanation.                                                 & 0.626                          & -0.111       & 0.636                           & -0.053        & 0.532                    & -0.294  & 0.675                    & -0.067  \\
2                     & Guaranteed features of the item but random capturing user preferences.            & 0.849                          & -0.100       & 0.787                           & -0.062        & 0.875                    & 0.000   & 0.923                    & 0.000   \\
3                     & Random features but guaranteed to capture user preferences.                          & 0.762                          & -0.064       & 0.852                           & -0.100        & 0.874                    & -0.083  & 0.928                    & -0.500  \\
4                     & Explanations generated using a state-of-the-art XRS method \cite{balloccu_post_2022} good at capturing features and user preferences. & 0.929                          & -0.100       & 0.935                           & 0.000         & 0.967                    & -0.138  & 0.983                    & 0.069   \\ \bottomrule
\end{tabular}
\label{tab:results1}
\vspace{-10pt}
\end{table*}

For setting different quality levels of explanations, we made four cases depending on whether Fidelity and Attunement are present.
In the first case, explanations did not exhibited Fidelity or Attunement.
While recommendations are generated using the baseline method \cite{balloccu_post_2022}, features are randomly selected and presented as explanations.
In the second case, explanations exhibited Fidelity but not Attunement.
Here, user preferences are not considered. 
Instead, explanations include features solely related to the recommended item.
In the third case, it was the opposite of the second case. Explanation exhibited Attunement but not Fidelity.
In the final case, we adopted the above mentioned baseline model to generate explanations exhibiting both Fidelity and Attunement.
This method leverages reinforcement learning-based path reasoning to identify the most suitable item for the user and provides the user-item path as an explanation.
The explanations incorporate features relevant to the recommended item while also capturing the user’s preferences.

For each case, we generated 30 (recommended item, feature) pairs.
To determine the Fidelity of each pair, if the feature in the provided pair is an actual attribute of the item, it is marked as true; otherwise, it is marked as false.
Regarding Attunement decisions, it is necessary to identify the features the user likes and dislikes.
These preferences were derived based on the user’s past ratings of items.
If a user has given a rating of 3 stars or higher to a specific item, the features associated with that item were considered user-liked features.
Conversely, if the user has rated an item below 3 stars, its features were regarded as user-disliked features.

\subsection{Results}

The explanation qualities calculated by our proposed metric for each case described above are presented in Table \ref{tab:results1}.
The sensitivity values ($A'$s) for Fidelity and Attunement for each case showed that the values increased proportionally with the quality settings from the cases.
Similarly, the $A'$ for Veracity tended to increase as explanations' quality improved.
In both the restrictive and permissive versions of Veracity, the trends of $A'$s are similar. 
However the values for the permissive version's were consistently higher (as indicated by a paired t-test; $t(119)=2.4511$, $p = 0.0142$) than the restrictive one.
This supports our hypothesis that the permissive setting, by allowing more flexible evaluation criteria, would yield higher sensitivity in measuring Veracity.
In terms of $B''_D$, all the values were shown to be around 0, indicating that the decision making process was not biased. 



\section{Discussion}
\label{sec:dis}

This research introduced an objective metric for quantifying the quality of explanations and attempted to validate its ability to distinguish between different performance conditions.
Our metric focused on Veracity, a factor that evaluates explanations from the perspective of information quality.
Veracity was analyzed along two dimensions: Fidelity, which assesses whether the explanation conveys accurate information; and Attunement, which measures whether it captures the preferences of the target user for whom the recommendation is made.
Using SDT, we defined decision outcomes for these two sub-factors and combined them to compute the final Veracity score using both restrictive and permissive methods.
To evaluate both version of our metric, we conducted experiments with four cases of explanations, assessing them to determine whether the results showed meaningful differences and significant insights.

Both versions of the metric (Veracity's sensitivity; $A'$) increased as the performance of the overall explanation increased across the Fidelity and Attunement dimensions: moving from effectively no sensitivity in Case 1, to increased values in Cases 2 and 3, to nearly perfect sensitivity in Case 4. 
The permissive version of Veracity's sensitivity appeared to consistently produce higher values than the sensitivities seen for Fidelity, Attunement, or restrictive Veracity. 
This likely makes the permissive version less useful than the restrictive one, as the sensitivity for restrictive Veracity exhibited clear differentiation between the values seen for Case 4 and those for Cases 2 and 3. This was lacking in the permissive version. 
Specifically, the permissive Veracity sensitivity measure rated explanation performance when exclusively either Fidelity or Attunement were high (Cases 2 and 3) as being comparable to situations where both were high (Case 4). 
If you accept our argument that both Fidelity and Attunement are important for explanation Veracity, then our recommendation would be to use the the restrictive version moving forward. 

All the bias ($B''_D$) values seen across the measures were close to 0, suggesting no bias. This makes sense given that our experiment did not attempt to set judgment critieria in a way that would bias results. Thus, this results provides confirmation that the bias measures are performing as intended. 

To the best of our knowledge, this study is the first to objectively examine the Veracity of an XRS explanation as a critical consideration. 
We believe this a major contribution, with the multidimensional nature of the Veracity measure potentially offering diagnostics for via the Fidelity and Attunement dimensions. 
The apparent success of the restrictive sensitivity for Veracity as a metric suggests multiple avenues of future research. 

This work did not vary the criterion threshold used in the judgments the XRS made in relation to Fidelity or Attunement. Future work could investigate how $B''_D$ values for Veracity change in response to such variation.

The restrictive and permissive approaches to computing Veracity's outcomes clearly impacted its sensitivity. 
There are other methods that could be used for computing these outcomes. 
For example, a balanced method could potentially split outcomes between the inconsistent versions (e.g., a Fidelity H and an Attunement FA and would be counted as 0.5 H and 0.5 FA outcomes for Veracity). 
Alternatively, Fidelity and Attunement outcomes could have different implications for different applications. This might suggest some form of weighting when synthesizing these into Veracity outcomes. 
Future work should explore these different options for computing Veracity outcomes, identify applications where different variations would be appropriate, and evaluate how they impact Veracity's sensitivity and bias measures. 

Finally, if a user notices that a an explanation provided by an XRS is not veracious, this will likely impact the user's opinion of that system. 
Future work should investigate how variation in Veracity (and its Fidelity and Attunement dimensions) impact human subjective ratings for the dimensions from \cref{tab:factors}. If there is a strong correlation, the measures introduced here could potentially be used instead of user studies to evaluate different dimensions of XRS explanations.

We can regard Veracity as a measure of the strength of the relationship between a target user and an item based on its features.
Thus, it could be used as an effective means of determining weights in XRS systems that are based on graphs or knowledge graphs. 
Alternatively, we can directly update existing item and feature representations to reflect user-specific preferences.
As part of our future work, we plan to develop a human-in-the-work framework that collects user feedback on the XRS’s results, calculates Fidelity, Attunement, and Veracity, and integrates this information into the recommendation backbone to generate improved results. 

\bibliographystyle{IEEEtranN}
\bibliography{IEEEabrv,references}


\end{document}